\documentclass[conference]{IEEEtran}
\IEEEoverridecommandlockouts
\usepackage{cite}
\usepackage{arydshln}
\usepackage{multirow}
\usepackage{tabularx}
\usepackage{amsmath,amssymb,amsfonts}
\usepackage{algorithmic}
\usepackage{booktabs}
\usepackage{graphicx}
\usepackage{textcomp}
\usepackage{xcolor}
\usepackage{graphicx}
\def\BibTeX{{\rm B\kern-.05em{\sc i\kern-.025em b}\kern-.08em
    T\kern-.1667em\lower.7ex\hbox{E}\kern-.125emX}}

\makeatletter
\def\adl@drawiv#1#2#3{%
        \hskip.5\tabcolsep
        \xleaders#3{#2.5\@tempdimb #1{1}#2.5\@tempdimb}%
                #2\z@ plus1fil minus1fil\relax
        \hskip.5\tabcolsep}
\newcommand{\cdashlinelr}[1]{%
  \noalign{\vskip\aboverulesep
           \global\let\@dashdrawstore\adl@draw
           \global\let\adl@draw\adl@drawiv}
  \cdashline{#1}
  \noalign{\global\let\adl@draw\@dashdrawstore
           \vskip\belowrulesep}}
\makeatother

\begin{document}

\title{Who Are They to Each Other? Multi-Agent Reasoning for Speaker Relationship Inference\\
}

\author{ \IEEEauthorblockN{ Yaohan Guan, Yen-Ju Lu, Yuzhe Wang, Junhyeok Lee, Jes\'us Villalba, \\ 
Laureano Moro Velazquez, Thomas Thebaud, Najim Dehak } \IEEEauthorblockA{ \textit{Center for Language and Speech Processing} \\ \textit{Department of Electrical and Computer Engineering} \\ \textit{Johns Hopkins University} \\ Baltimore, MD, USA \\ \{yguan19, ylu125, ywang792, jvillal7, laureano, tthebau1, ndehak3\}@jhu.edu } }


\maketitle

\begin{abstract}

Inferring speaker relationships from spoken conversations is an important step towards socially aware speech understanding. However, this task remains underexplored, and supervised modeling is costly to train and scale. At the same time, existing inference-time LLM approaches provide limited structure for handling subtle, distributed, and multimodal relational cues that may support multiple plausible interpretations. To address these limitations, we introduce a training-free multi-agent reasoning framework that organizes inference through structured interaction among LLM agents, allowing relationship judgments to be proposed, challenged, and adjudicated
without task-specific training. We instantiate this framework with two complementary designs. 
We propose Multi-Role Multi-Agent Debate as a task-specific adaptation of standard multi-agent debate for speaker relationship inference, assigning agents complementary roles or social-theory-grounded perspectives rather than a single undifferentiated viewpoint. In contrast, we introduce Multi-Agent Compete, a competition-based protocol that compares agent judgments through pairwise adjudication, eliminates weaker candidates, and retains the most defensible one. We evaluate these methods on the Seamless Interaction dataset across different modality settings, covering both binary familiar-versus-stranger classification and fine-grained relationship-detail prediction. Results suggest that they improve over zero-shot and existing multi-agent baselines in most cases. Human evaluation further suggests that speaker relationship inference is challenging even for people. LLM methods can sometimes outperform human annotators in text-included settings but are less competitive in the audio setting. Together, these findings suggest that relationship inference benefits from structured inference-time interaction among agents, while acoustic cues are not yet fully captured by current models.

\end{abstract}


\begin{figure*}[t] 
    \centering
    \includegraphics[width=\textwidth]{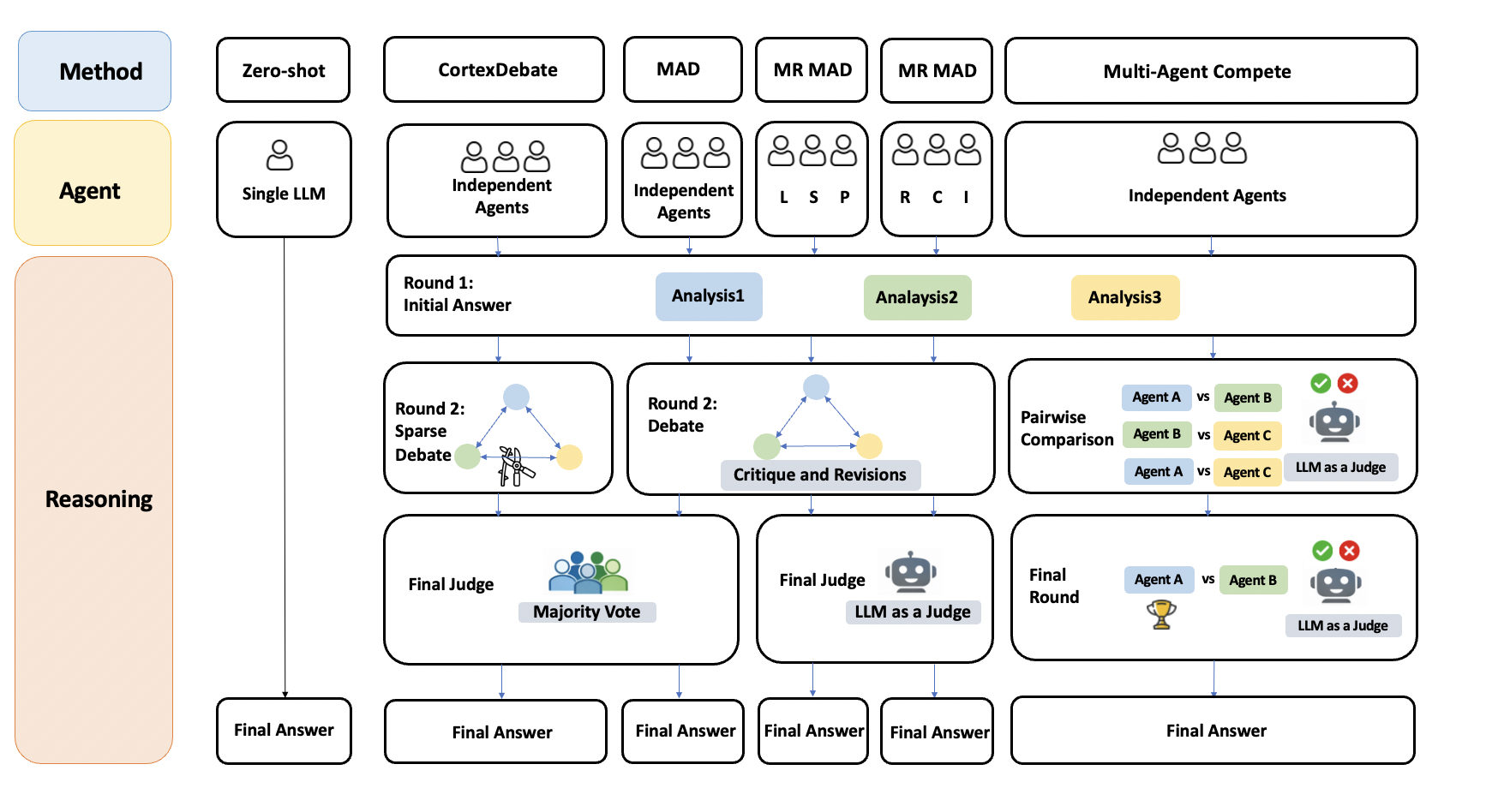}
    \vspace{-12mm}
\caption{Overview of six inference-time reasoning frameworks compared in this work. 
}
    \label{fig:Method}
    \vspace{-4mm}
\end{figure*}
\section{Introduction}


Human conversations carry rich social information beyond their literal content. The way people speak to each other—what they say, how they say it, and how they coordinate—reflects the nature of their relationship. Understanding these relational signals is central to building socially aware systems that can interpret not just what is said, but the social context in which it is said \cite{yang2025socially}. Speaker relationship inference—determining how two speakers are socially connected from their conversation—is one instantiation of this challenge.

Despite its importance, speaker relationship inference remains underexplored. Prior work has largely focused on text-based or feature-engineered settings, relying on lexical patterns, topic modeling, and dialogue structure to distinguish broad relationship categories \cite{stark2012hello, tigunova2021pride}. Some earlier work incorporated acoustic features \cite{yella2014inferring, katerenchuk2014your}, but relied on hand-crafted feature extractors and covered only coarse-grained distinctions. Spoken conversation datasets with fine-grained relationship annotations remain scarce, leaving the multimodal nature of spoken relationship cues—where prosody, speaking style, and interactional dynamics carry information text alone cannot capture—largely understudied.

Speaker relationship inference is also inherently difficult. Relationship cues are subtle and distributed across turns rather than concentrated in a single utterance, and the same surface signal—a friendly tone, a casual register—can be consistent with multiple relationship types. In speech, evidence may appear across both linguistic and acoustic channels, requiring models to integrate complementary modalities. Supervised approaches face additional barriers: socially grounded annotations are costly to collect, and trained models may not generalize well across relationship categories or input modalities.



Given this difficulty, two questions naturally arise: how well do current Large Language Models (LLMs) and Multimodal LLMs perform on this task, and does performance vary across modalities? And how can inference-time strategies improve performance without additional training? Because relational cues often support multiple interpretations, answering the second question requires methods that can compare competing hypotheses. Multi-Agent Debate \cite{du2024improving}, where LLM agents iteratively propose, critique, and revise answers before producing a final decision, offers a natural fit. Recent work has shown its effectiveness for logical reasoning and factuality, yet it has rarely been studied for this kind of socially grounded task. We therefore propose a new multi-agent reasoning framework; unlike prior multi-agent debate approaches that employ homogeneous reasoning agents, ours explicitly structures inference around complementary social perspectives and competition-based hypothesis selection, intended to better capture the ambiguity and distributed evidence characteristic of speaker relationship inference. 

Our framework includes two complementary designs. The first, Multi-Role Multi-Agent Debate (MRMAD), adapts standard multi-agent debate by assigning agents complementary analytic roles or social theories rather than a single undifferentiated viewpoint. The second, Multi-Agent Compete (MAC), introduces a competition-based protocol in which agents independently propose relationship hypotheses that are then compared through pairwise adjudication, eliminating weaker candidates and retaining the most defensible interpretation. We test the framework on the Seamless Interaction dataset, a naturalistic spoken conversation corpus supporting both binary familiar-versus-stranger classification and fine-grained relationship-detail prediction, evaluating it across text only, audio only, and audio plus text to assess its applicability across modalities. Experiments show that our design improves over zero-shot and standard multi-agent baselines in most cases. Human evaluation further reveals that speaker relationship inference is challenging even for people; LLM-based methods can sometimes exceed human performance in text-based settings but are less competitive when audio is the primary modality. These suggest two broader conclusions: relationship inference benefits from structured inference-time interaction among agents, and current multimodal LLMs do not yet fully exploit acoustic relational cues for this task.

\section{Related work}
\subsection{Speaker Relationship Inference}
Prior work has explored how interpersonal relationships can be inferred from conversational interactions in both text and speech settings. This work\cite{stark2012hello} investigated language-use features, including entropy-based measures, hand-crafted lexical categories from Linguistic Inquiry and Word Count (LIWC), and topics learned with latent Dirichlet allocation (LDA), finding that posteriors over LDA topics consistently outperformed LIWC-based or language-use features for distinguishing conversation types. PRIDE \cite{tigunova2021pride} is a neural multi-label classifier based on BERT and a Transformer for constructing conversation representations, leveraging dialogue structure and augmenting it with external knowledge about speaker attributes and conversation style. In speech-based settings, authors\cite{yella2014inferring} trained a boosting-based classifier on conversational and acoustic features to classify phone calls by the social relationship between speakers, while authors in\cite{katerenchuk2014your} used lexical and acoustic features from dialogue to classify interpersonal relationships such as family versus friends.

\subsection{Multi-Agent Reasoning}
Recent work has explored multi-agent interaction as an inference-time strategy for improving the reasoning ability of LLMs. A representative work is Multi-Agent Debate (MAD), where multiple LLM agents iteratively propose, critique, and revise answers before producing a final decision \cite{du2024improving}. This paradigm has been shown to improve factuality and reasoning performance, suggesting that structured interaction among agents can help expose errors and refine intermediate judgments. Subsequent studies have extended this idea by modifying how agents communicate and how their outputs are aggregated. Authors in \cite{wu2025identifying} apply multi-agent reasoning to conversational social reasoning by identifying power relations through rebuttal, scoring, and factor-graph-based inference. Other work improves the debate process itself: CortexDebate constructs a sparse debating graph so that each agent interacts only with useful peers, mitigating the overlength and overconfidence issues of standard MAD, while MAD-Logic introduces debate in both natural language and symbolic representations to strengthen logical reasoning \cite{yang2026mad}.


Together, these two lines of work leave a gap at their intersection. Speaker relationship inference has relied on coarse-grained categories and statistical or feature-based models, with little known about how LLMs perform. Multi-agent reasoning, in turn, has improved logical and factual reasoning but has not been applied to relationship inference, nor designed with roles suited to weighing ambiguous, distributed social cues. To address this gap, we introduce a multi-agent reasoning framework, organizing inference through structured interaction among LLM agents that allows relationship judgments to be proposed, challenged, and adjudicated without training.
\section{Dataset}
We evaluate our methods on the Seamless Interaction dataset~\cite{agrawal2025seamless}, a large-scale collection of in-person dyadic interactions designed for social AI research. 
The full dataset contains over 4,000 hours of interactions across \textit{Naturalistic} and \textit{Improvised} settings. 
We focus on the Naturalistic subset, where participants interact with people they may have known previously, making it more suitable for studying naturally occurring speaker relationships than the Improvised setting, where actors perform assigned scenarios. The Naturalistic subset contains both coarse-grained labels (Familiar, Stranger, Unknown) and fine-grained relationship detail labels. Familiar relationships are further categorized
into nine categories: friends, coworkers, family-generic, familiar-generic, dating/spouse/romantic partner, classmates, siblings, parent/child,
neighbors, and roommates, though the test set only covers
friends, family, romantic partner, and coworkers.

We constructed the evaluation set by applying three filtering criteria. First, we excluded 56 dialogues labeled \textit{Unknown}, restricting evaluation to cases with unambiguous ground-truth relationship labels. Second, we removed 44 dialogues with empty transcripts, which indicate failed or unusable transcriptions. We also excluded the familiar-generic label. The remaining relationship categories can each be grounded in a well-defined criterion: kinship ties (e.g., family-generic, siblings, parent/child), structural or contextual bonds (e.g., coworkers, roommates, neighbors), or clearly specified social relations (e.g., friends, dating/spouse/romantic partner). In contrast, familiar-generic lacks any comparable defining criterion—it functions as a residual category for dyads that are acquainted but do not fit the other classes, rather than corresponding to a coherent relational construct. We therefore exclude it to ensure that all retained labels correspond to clearly interpretable relationship categories. After filtering, the final test set contains 607 dialogues, including 441 familiar (72.7\%) and 166 stranger (27.3\%) samples. The familiar dialogues span four relationship subtypes: friends ($n=273$), family ($n=102$), romantic partners ($n=51$), and coworkers ($n=15$). For fair comparison, all text, audio, and audio+text modality results are reported on this same filtered test set.



\section{Methods}

The task is to first predict whether the two speakers in a conversation are familiar or strangers. If familiar, the model further classifies the specific relationship into one of nine categories(after excluding familiar-generic). We retain these nine categories rather than the four available in the test set, since the train and development sets are annotated at this finer granularity—also leaving open compatibility with other datasets that may adopt similarly fine-grained relationship labels. We include 6 methods: three baselines (Zero-shot, Multi-Agent Debate, and CortexDebate) and our methods (two variations of Multi-Role Multi-Agent Debate, and Multi-Agent Compete).




\subsubsection{Zero-shot}
We directly prompt the model with the interaction to get the final predictions. This would be the equivalent of a Single Agent Single Role scenario.

\subsubsection{Multi-Agent Debate}
We follow the Multi-Agent Debate framework~\cite{du2024improving}, where three agents using the same base model first independently predict the relationship and its details. In the later two rounds, agents revise their responses based on the others' opinions. All agents receive the same prompt in each round, with final predictions determined by majority vote.

\subsubsection{CortexDebate}
CortexDebate~\cite{sun2025cortexdebate} extends Multi-Agent Debate with an adaptive communication structure: after an initial round where all three agents respond, a McKinsey-based Debate Matter (MDM) module estimates pairwise trustworthiness, then builds a sparse debate graph so agents exchange information only with the most helpful partners. Debate ends early on consensus or at the round limit (n=2), with the final prediction by majority vote.

\subsubsection{Multi-Role Multi-Agent Debate}

As shown in Figure~\ref{fig:Method}, we propose a Multi-Role Multi-Agent Debate framework that extends standard MAD by assigning agents complementary roles. Inspired by multi-head attention, where different heads attend to different input subspaces, each agent is prompted to attend to a different aspect of the conversation. We instantiate this idea in two configurations.

The first variant instantiates three agents with complementary analytical roles: a \textbf{Linguist}, a \textbf{Sociologist}, and a \textbf{Psychologist}. 
We refer to this role configuration as \textbf{LSP}, and denote the resulting framework as \textbf{Multi-Role (LSP) Multi-Agent Debate}. These perspectives were selected because they capture complementary aspects of interpersonal relationships that have been extensively studied in linguistics, psychology, and sociology, providing diverse yet theoretically grounded views of conversational social dynamics. The linguist attends to lexical, syntactic and semantic levels; the psychologist focuses on socio-emotional dynamics such as empathy, alignment, and rapport; and the sociologist examines social roles, norms, authority, deference, and status cues.

The second configuration, \textbf{Multi-Role (RCI) Multi-Agent Debate}, assigns agents theory-grounded roles and structures debate through quantified relational dimensions. The first agent is based on \textbf{Relational Models Theory}~\cite{fiske1990relativity,fiske1992four}, covering communal sharing, authority ranking, equality matching, and market pricing. The second follows \textbf{Communal/Exchange Theory}~\cite{clark1979interpersonal}, focusing on communal and exchange dimensions. The third follows \textbf{Interpersonal Circumplex Theory}~\cite{pincus2003interpersonal,wright2023contemporary}, focusing on agency and communion. Each agent scores the conversation along its theory-derived dimensions. These dimension-level scores serve as the medium of debate: rather than exchanging vague natural-language impressions, agents challenge and revise each other's scored assessments, grounding the discussion in explicit and quantifiable relational judgments before a final prediction is reached.

For both methods, agents would have two rounds of debate after the initial round. The final judge is LLM-as-a-judge using the same base model as the agents.



\subsubsection{Multi-Agent Compete}Different from the debate-based framework, where agents collaborate to some extent by helping one another examine possible flaws, the Multi-Agent Compete method puts them in direct competition, creating stronger incentive pressure for each agent to produce the best possible answer. Concretely, as shown in Figure \ref{fig:Method}, three agents first produce their initial submissions independently, with the prompt explicitly stating that the interaction is an elimination tournament. In the second round, an LLM judge conducts all pairwise comparisons among the three agent outputs (A vs. B, B vs. C, and A vs. C) and selects the stronger answer in each matchup based on overall correctness, rule consistency, and evidence quality. One agent is then eliminated according to the pairwise outcomes, and this elimination is broadcast to the remaining agents. In the final stage, the two surviving agents revise their answers after seeing the tournament broadcast and the summary of earlier pairwise results, and the judge performs a final head-to-head comparison to select the winner. The final prediction is taken from the winning agent’s revised output.

\section{Experimental settings and Results}

\subsection{Experimental Settings}
We evaluate all methods on the same filtered test set across three modalities: text, audio, and audio+text. For text, we use GPT-5-mini; for audio and audio+text, we compare closed- and open-source models using GPT-audio-1.5 and Qwen2.5-Omni-7B, respectively, with temperature set to 1 for GPT-5-mini and GPT-audio-1.5. We report accuracy and macro F1 for both binary and multi-class classification, where multi-class metrics are computed only on Familiar samples ($n=441$). It is important to note that the relationship categories are highly imbalanced, with some classes (e.g., coworkers) containing relatively few examples, so per-category metrics should be interpreted cautiously while the overall binary and multi-class metrics provide a more reliable estimate of system performance. Given this imbalance, macro F1 should be emphasized over accuracy, as it better reflects performance across underrepresented categories.
\subsection{Result Analysis}
\subsubsection{Text Modality}

\begin{table*}[t]
\centering
\vspace{-5mm}
\footnotesize
\caption{\small Classification results for Binary ($n$=607), multi-class relationship-detail classification ($n$=441), and per-category performance for Friends ($n$=273), Family ($n$=102), Dating/Romantic ($n$=51), Coworkers ($n$=15). Bold and underlining indicate the best and second-best results, respectively, within each column of each modality--model block.}
\vspace{-3mm}
\label{tab:merged_results}
\resizebox{\textwidth}{!}{%
\begin{tabular}{@{}lllcccc|cccccccc@{}}
\toprule
& & & \multicolumn{2}{c}{Binary} & \multicolumn{2}{c|}{Multi-class} & \multicolumn{2}{c}{Friends} & \multicolumn{2}{c}{Family} & \multicolumn{2}{c}{Dating/Romantic} & \multicolumn{2}{c}{Coworkers}\\
\cmidrule(lr){4-5}\cmidrule(lr){6-7}\cmidrule(lr){8-9}\cmidrule(lr){10-11}\cmidrule(lr){12-13}\cmidrule(lr){14-15}
Modality & Model & Method & Macro F1 & Acc. & Macro F1 & Acc. & F1 & Acc. & F1 & Acc. & F1 & Acc. & F1 & Acc. \\
\midrule
\multirow{6}{*}{Text}
& \multirow{6}{*}{GPT-5 mini} & Zero-Shot            & 60.4\% & 67.7\%& 11.3\% & 42.6\%& 67.9\% & 62.6\%& 3.8\% & 2.0\%& \textbf{28.0\%} & \textbf{25.5\%}& 12.9\% & 13.3\% \\
& & Multi-Agent Debate   & 61.2\% & 68.0\%& 11.1\% & 40.1\%& 66.9\% & 58.6\%& 3.8\% & 2.0\%& \textbf{28.0\%} & \textbf{25.5\%}& 12.5\% & 13.3\% \\
& & CortexDebate         & \underline{62.0\%} & 67.6\%& 11.6\% & 41.5\%& 67.6\% & 60.1\%& \underline{7.3\%} & \underline{3.9\%}& \textbf{28.0\%} & \textbf{25.5\%}& 13.3\% & 13.3\% \\
\cdashlinelr{3-15}
& & \textit{Multi-Role (LSP) MAD (Ours)}& 60.3\% & 67.7\%& \underline{12.3\%} & 44.7\%& 69.2\% & \underline{64.1\%}& \textbf{19.1\%} & \textbf{10.8\%}& 23.1\% & 17.6\%& 11.8\% & 13.3\% \\
& & \textit{Multi-Role (RCI) MAD (Ours)}& 60.5\% & \underline{69.9\%}& 12.0\% & \underline{45.8\%}& \underline{70.3\%} & \textbf{67.8\%}& 3.8\% & 2.0\%& \underline{26.7\%} & \underline{23.5\%} & \underline{19.4\%} & \underline{20.0\%} \\
& & \textit{Multi-Agent Compete (Ours)} & \textbf{62.5\%} & \textbf{71.2\%}& \textbf{12.4\%} & \textbf{46.0\%}& \textbf{71.2\%} & \textbf{67.8\%}& 5.6\% & 2.9\%& 24.2\% & 21.6\% & \textbf{22.9\%} & \textbf{26.7\%} \\
\midrule
\midrule
\multirow{14}{*}{Audio+Text}
& \multirow{6}{*}{GPT-audio-1.5} & Zero-Shot            & 53.8\% & 62.3\% & 29.4\% & 42.0\% & 67.1\% & 61.9\%& 7.5\% & 3.9\%& 20.3\% & 15.7\%& 22.9\% & \underline{26.7\%} \\
&  & Multi-Agent Debate   & 63.2\% & 73.6\% & 33.1\% & \underline{51.7\%} & \underline{73.7\%} & \underline{75.1\%}& 10.8\% & 5.9\%& 30.8\% & \underline{27.5\%}& 17.1\% & 20.0\% \\
&  & CortexDebate         & 63.2\% & \textbf{75.0\%} & 33.6\% & \textbf{53.3\%} & \textbf{74.7\%} & \textbf{77.7\%}& 10.8\% & 5.9\%& \textbf{31.1\%} & \underline{27.5\%}& 17.7\% & 20.0\% \\
\cdashlinelr{3-15}
& & \textit{Multi-Role (LSP) MAD (Ours)} & \underline{64.6\%} & 71.0\% & \textbf{36.0\%} & 49.2\% & 71.3\% & 69.2\%& \textbf{26.6\%} & \textbf{16.7\%}& 18.7\% & 13.7\%& \textbf{27.6\%} & \underline{26.7\%} \\
&  & \textit{Multi-Role (RCI) MAD (Ours)} & 59.8\% & 73.2\% & \underline{33.9\%} & 51.3\% & 72.9\% & 74.0\%& \underline{12.3\%} & \underline{6.9\%}& 27.3\% & 23.5\%& \underline{23.3\%} & \textbf{33.3\%} \\
&  & \textit{Multi-Agent Compete (Ours)}  & \textbf{66.5\%} & \underline{74.1\%} & 33.0\% & 48.8\% & 71.4\% & 70.0\%& 10.9\% & 5.9\%& \underline{30.9\%} & \textbf{29.4\%}& 18.8\% & 20.0\% \\
\cmidrule(lr){2-15}

& \multirow{6}{*}{Qwen2.5-Omni-7B} & Zero-Shot            & 41.4\% & 70.2\% & 28.9\% & 57.1\% & 74.4\% & 86.1\%& 21.3\% & \underline{12.8\%}& 0.0\% & 0.0\%& 20.0\% & 26.7\% \\
&  & Multi-Agent Debate   & 42.4\% & \underline{72.0\%} & 27.5\% & \underline{60.1\%} & \underline{76.0\%} & \underline{92.3\%}& 17.0\% & 9.8\%& 0.0\% & 0.0\%& 17.1\% & 20.0\% \\
&  & CortexDebate         & 41.9\% & \underline{72.0\%} & 30.0\% & \textbf{61.2\%} & \textbf{77.1\%} & \textbf{93.0\%}& 20.0\% & 11.8\%& 0.0\% & 0.0\%& 22.9\% & 26.7\% \\
\cdashlinelr{3-15}
&  & \textit{Multi-Role (LSP) MAD (Ours)} & 43.5\% & 70.7\% & 30.3\% & 54.7\% & 72.6\% & 81.0\%& 21.3\% & \underline{12.8\%}& 3.4\% & 2.0\%& \underline{24.0\%} & \underline{40.0\%} \\
&  & \textit{Multi-Role (RCI) MAD (Ours)} & \textbf{45.3\%} & \textbf{72.5\%} & \underline{33.2\%} & 57.1\% & 74.8\% & 83.5\%& \textbf{22.8\%} & \textbf{13.7\%}& \textbf{12.3\%} & \textbf{7.8\%}& 23.1\% & \underline{40.0\%} \\
&  & \textit{Multi-Agent Compete (Ours)}  & \underline{44.2\%} & 70.0\% & \textbf{33.4\%} & 54.0\% & 71.7\% & 78.4\%& \underline{21.7\%} & \underline{12.8\%}& \underline{6.1\%} & \underline{3.9\%}& \textbf{34.0\%} & \textbf{60.0\%} \\

\midrule
\midrule
\multirow{12}{*}{Audio}


& \multirow{6}{*}{GPT-audio-1.5} & Zero-Shot            & 44.6\% & 69.9\% & 35.8\% & 54.0\% & 74.6\% & 76.6\%& \underline{17.7\%} & \underline{9.8\%}& \textbf{31.1\%} & \textbf{27.5\%}& 19.6\% & \underline{33.3\%} \\
&  & Multi-Agent Debate   & \textbf{50.8\%} & \textbf{74.0\%} & 36.7\% & \underline{59.0\%} & \underline{78.5\%} & \underline{85.7\%}& 14.3\% & 7.8\%& 27.7\% & \underline{25.5\%}& \underline{26.3\%} & \underline{33.3\%} \\
&  & CortexDebate         & 45.7\% & 71.8\% & 36.6\% & \textbf{59.6\%} & \textbf{79.5\%} & \textbf{87.2\%}& 12.3\% & 6.9\%& \underline{29.8\%} & \textbf{27.5\%}& 25.0\% & 26.7\% \\
\cdashlinelr{3-15}
&  & \textit{Multi-Role (LSP) MAD (Ours)} & 47.0\% & 70.2\% & 33.6\% & 54.9\% & 72.8\% & 76.6\%& \textbf{32.7\%} & \textbf{24.5\%}& 17.7\% & 11.8\%& 11.4\% & 13.3\% \\
& & \textit{Multi-Role (RCI) MAD (Ours)} & 48.6\% & \underline{72.8\%} & \textbf{37.2\%} & 58.3\% & 77.5\% & 83.9\%& 17.5\% & \underline{9.8\%}& 28.6\% & \underline{25.5\%}& 25.0\% & \underline{33.3\%} \\
&  & \textit{Multi-Agent Compete (Ours)}  & \underline{49.8\%} & 70.0\% & \underline{36.9\%} & 55.3\% & 75.6\% & 79.9\%& 10.8\% & 5.9\%& \underline{29.8\%} & \textbf{27.5\%}& \textbf{31.6\%} & \textbf{40.0\%} \\
\cmidrule(lr){2-15}

& \multirow{6}{*}{Qwen2.5-Omni-7B} & Zero-Shot            & 40.7\% & 66.1\% & 31.5\% & 54.4\% & 72.9\% & 79.5\%& \textbf{26.4\%} & \textbf{16.7\%}& \underline{3.6\%} & \underline{2.0\%}& 23.3\% & 33.3\% \\
&  & Multi-Agent Debate   & 42.5\% & \textbf{72.3\%} & \underline{32.1\%} & \underline{58.7\%} & \underline{74.3\%} & \underline{86.8\%}& \underline{23.8\%} & \underline{14.7\%}& 0.0\% & 0.0\%& 30.4\% & \underline{46.7\%} \\
&  & CortexDebate         & 42.4\% & \underline{71.8\%} & \textbf{32.7\%} & \textbf{59.2\%} & \textbf{74.8\%} & \textbf{87.9\%}& 22.6\% & 13.7\%& 0.0\% & 0.0\%& \textbf{33.3\%} & \underline{46.7\%} \\
\cdashlinelr{3-15}
&  & \textit{Multi-Role (LSP) MAD (Ours)} & \underline{49.3\%} & 60.8\% & 28.9\% & 43.8\% & 66.0\% & 64.5\%& 15.7\% & 8.8\%& 3.5\% & \underline{2.0\%}& 30.4\% & \underline{46.7\%} \\
&  & \textit{Multi-Role (RCI) MAD (Ours)} & \textbf{50.3\%} & 67.2\% & 28.8\% & 47.9\% & 67.5\% & 71.1\%& 15.3\% & 8.8\%& \textbf{6.5\%} & \textbf{3.9\%}& 26.1\% & 40.0\% \\
&  & \textit{Multi-Agent Compete (Ours)}  & 46.2\% & 68.9\% & 29.3\% & 52.2\% & 71.0\% & 78.0\%& 15.4\% & 8.8\%& 0.0\% & 0.0\%& \underline{30.8\%} & \textbf{53.3\%} \\
\bottomrule
\end{tabular}%
}
\vspace{-5mm}
\end{table*}

\begin{table}[t]
\centering
\small

\caption{Accuracy of zero-shot baselines with different models on relationship and relationship-detail prediction. GPT-5 mini achieves the highest accuracy on both tasks.}
\vspace{-3mm}
\begin{tabular}{@{}lcc@{}}
\toprule
Method & Relationship Acc. & Detail Acc. \\
\midrule
Zero-shot GPT-5 mini & \textbf{67.7\%} & \textbf{42.6\%} \\
Zero-shot GPT-5.4 & 65.7\% & 41.4\% \\
Zero-shot GPT-5.4 (low) & 64.4\% & 32.8\% \\
Zero-shot GPT-5.4 (xhigh) & 62.2\% & 25.5\% \\
\bottomrule
\vspace{-8mm}
\end{tabular}

\label{tab:filtered_zero_shot_overall}
\end{table}

Table~\ref{tab:merged_results} compares our multi-agent methods with zero-shot and two debate-based baselines. For binary classification, \textit{Multi-Agent Compete} achieves the best result among our three proposed methods. This improves  macro F1 by 0.5 points over the strongest baseline, \textit{CortexDebate}, and accuracy by 3.2 points over the strongest baseline, \textit{Multi-Agent Debate}. The advantage of structured agent interaction becomes more evident in relationship-detail prediction, where \textit{Multi-Agent Compete} again performs best among our methods, getting improvements of 0.8 points in macro F1 and 3.4 points in accuracy over the strongest baseline. All three proposed methods outperform every baseline on multi-class macro F1 and accuracy. The per-category results in Table~\ref{tab:merged_results} further suggest that the gains are concentrated primarily in the Friends and Coworkers categories, where our multi-agent methods—particularly \textit{Multi-Agent Compete}—consistently outperform all baselines in both accuracy and F1. However, this improvement is not uniform across relationship types. On Dating/Romantic, all baselines perform identically, while every multi-agent method underperforms this baseline.


We also examine whether stronger models or more intensive reasoning directly improve speaker relationship inference. As shown in Table~\ref{tab:filtered_zero_shot_overall}, interestingly, GPT-5 mini consistently outperforms the GPT-5.4 variants despite their increased reasoning capability. We hypothesize that speaker relationship inference depends less on long-chain deductive reasoning and more on effectively integrating subtle conversational evidence distributed across multiple dialogue turns. This suggests that structuring the reasoning process may be more beneficial than simply increasing reasoning depth.
\subsubsection{Audio+Text Modality}
The results suggest that our methods continue to have an advantage on macro F1 for both binary and multi-class classification, which better reflects performance across imbalanced categories. On GPT-audio-1.5, Multi-Agent Compete achieves the best binary macro F1, outperforming the strongest baseline by 3.3 points, while Multi-Role (LSP) leads on multi-class macro F1, 2.4 points above CortexDebate. On Qwen2.5-Omni-7B, Multi-Role (RCI) achieves the best binary accuracy and also surpasses the best baseline by 2.9 points in binary macro F1; Multi-Agent Compete also leads on multi-class macro F1. This indicates that when both linguistic and acoustic cues are available, our structured multi-agent reasoning not only remains effective but also yields larger gains on macro F1.

\subsubsection{Audio Modality}
In this setting, our methods show a mix of gains and modest declines relative to the baselines, though the advantage on macro F1 is largely preserved. For instance, under GPT-audio-1.5, Multi-Role (RCI) MAD achieves the best multi-class macro F1 among all methods, and under Qwen2.5-Omni-7B, it likewise achieves the best binary macro F1, an improvement of 7.8 points over the strongest baseline. At the same time, binary macro F1 under GPT-audio-1.5 and multi-class macro F1 under Qwen2.5-Omni-7B fall behind the best baselines. However, in binary macro F1 under GPT-audio-1.5, our Multi-Agent Compete still lands in second place. This suggests that in the audio-only setting, our multi-agent reasoning remains effective at balancing predictions across relationship categories, but is less able to help the models as consistently as it does under the text and audio+text modalities, where our methods occupy a larger share of the best- and second-best positions across metrics.

Taken together, our methods obtain the largest and consistent macro F1 gains in the Audio+Text modality, a smaller but still consistent gain in the Text modality, and a mixed picture in the Audio-only modality, with some metrics improving and others declining slightly. 
One possible explanation is that debate-style methods were usually designed around the kind of explicit, quotable evidence that is naturally available in text transcripts, but is harder to extract and argue over from audio alone. Without clear textual evidence to anchor their arguments, agents in the audio-only setting may have less to debate over, which would explain why the resulting gains are smaller and less consistent than in the other two modalities. The fact that Audio+Text yields the largest and most consistent improvements on macro-F1 might indicate that the two modalities together provide more information that our multi-agent framework can exploit than either modality alone. This is further supported by comparing the same model and method across the Audio and Audio+Text modalities directly. This indicates that adding text on top of audio still gives an overall gain for our multi-agent framework in most metrics.

Among our three proposed methods, Multi-Agent Compete performs best in the text modality, while Multi-Role (RCI) MAD and Multi-Agent Compete tend to be the stronger choices under audio and audio+text. One possible explanation for the effectiveness of Multi-Agent Compete is that it preserves multiple competing hypotheses throughout most of the inference process instead of encouraging early consensus. Because speaker relationship inference often admits several plausible interpretations, maintaining diverse hypotheses before the final adjudication may allow the system to better resolve ambiguous social cues than collaborative debate alone.

\section{Human Evaluation Study}
\subsection{Human Evaluation Setting}
To assess the difficulty of this task for humans, we conducted a human evaluation study examining how accurately humans can infer speaker relationships across different input modalities. We sampled 50 conversations from the filtered test set and recruited 27 native-English-speaking annotators via Prolific, randomly assigned to one of three modalities: \textit{text only}, \textit{audio only}, or \textit{text + audio}. To assess annotator reliability, 5 of the 50 conversations were selected as fully overlapped within each modality: the remaining 45 conversations were each annotated by three annotators, while each overlapped conversation was annotated by all nine annotators assigned to that modality. This yielded 180 ratings per modality—135 from non-overlapped conversations and 45 from overlapped conversations.

\begin{table}[t]
\centering
\footnotesize
\caption{Inter-annotator agreement (Krippendorff's $\alpha$) across
modalities, computed on binary and fine-grained labels.}
\vspace{-3mm}
\label{tab:iaa}
\begin{tabular}{@{}lcccc@{}}
\toprule
Modality & \#Annot. & \#Ratings & $\alpha_{\text{bin}}$ & $\alpha_{\text{multi}}$ \\
\midrule
Text         & 9 & 180 & 0.206 & 0.171 \\
Audio+Text   & 9 & 180 & 0.225 & 0.250 \\
Audio        & 9 & 180 & 0.340 & 0.289 \\
\bottomrule
\end{tabular}
\vspace{-5mm}
\end{table}

\begin{table}[t]
\centering
\footnotesize
\caption{Human and method performance on the 50-sample test set across three modalities. Human rows pool individual annotator votes.  Multi-class results are computed on the $n{=}37$ samples with \textit{familiar} ground truth. Models: GPT-5 mini for text, and GPT-audio-1.5 for the other two modalities.}
\vspace{-3mm}
\label{tab:modality-results}
\resizebox{\linewidth}{!}{%
\begin{tabular}{@{}llcccc@{}}
\toprule
Modality & Method & Acc$_{\text{bin}}$ & F1$_{\text{bin}}$ & Acc$_{\text{multi}}$ & F1$_{\text{multi}}$ \tabularnewline
\midrule
\multirow{7}{*}{Text}
& Human ($n{=}180$) & 70.6\% & 57.8\% & 29.6\% & 26.8\% \tabularnewline
& Zero-Shot & \textbf{76.0\%} & \textbf{68.8\%} & 48.7\% & 28.9\% \tabularnewline
& MAD & 68.0\% & 60.3\% & 43.2\% & 25.8\% \tabularnewline
& CortexDebate & 72.0\% & 63.6\% & 45.9\% & 26.8\% \tabularnewline
& MR (LSP) MAD & 74.0\% & 63.3\% & \textbf{51.4\%} & 27.1\% \tabularnewline
& MR (RCI) MAD & 70.0\% & 57.7\% & 48.7\% & \textbf{31.7\%} \tabularnewline
& MAC & \textbf{76.0\%} & 65.0\% & 43.2\% & 31.1\% \tabularnewline
\midrule
\multirow{7}{*}{Audio+Text}
& Human ($n{=}180$) & 72.2\% & 65.4\% & 31.9\% & 32.2\% \tabularnewline
& Zero-Shot & 70.0\% & 60.0\% & 40.5\% & 25.0\% \tabularnewline
& MAD & 78.0\% & \textbf{66.8\%} & 48.6\% & 31.7\% \tabularnewline
& CortexDebate & \textbf{80.0\%} & 62.8\% & \textbf{56.8\%} & 25.6\% \tabularnewline
& MR (LSP) MAD & 74.0\% & 65.3\% & 51.4\% & \textbf{44.0\%} \tabularnewline
& MR (RCI) MAD & 78.0\% & 61.0\% & 51.4\% & 31.2\% \tabularnewline
& MAC & 76.0\% & 66.1\% & 45.9\% & 32.0\% \tabularnewline
\midrule
\multirow{7}{*}{Audio}
& Human ($n{=}180$) & 67.8\% & \textbf{62.8\%} & 30.4\% & 32.1\%
\tabularnewline
& Zero-Shot & 74.0\% & 54.7\% & 54.1\% & 32.9\% \tabularnewline
& MAD & \textbf{78.0\%} & 56.9\% & \textbf{59.5\%} & \textbf{34.3\%} \tabularnewline
& CortexDebate & 74.0\% & 49.5\% & 51.4\% & 31.0\% \tabularnewline
& MR (LSP) MAD & 74.0\% & 56.0\% & 51.4\% & 22.5\% \tabularnewline
& MR (RCI) MAD & 74.0\% & 49.0\% & 54.1\% & 31.8\% \tabularnewline
& MAC & 74.0\% & 55.2\% & 51.4\% & 31.8\% \tabularnewline
\bottomrule
\end{tabular}}
\vspace{-5mm}
\end{table}

\subsection{Inter-Annotator Agreement}
We measure inter-annotator agreement using Krippendorff's $\alpha$ \cite{krippendorff2018content} as shown in Table~\ref{tab:iaa}. Overall, agreement is low across the board. Rather than a weakness of the evaluation itself, this highlights the intrinsic ambiguity of speaker relationship inference: many conversational cues admit multiple plausible social interpretations, making disagreement expected even among native speakers. This further motivates structured inference-time reasoning that explicitly compares competing hypotheses instead of committing to a single interpretation early on. Across modalities, agreement follows a consistent ordering on both binary and multi-class labels: Text yields the lowest $\alpha$, Audio the highest, with Audio+Text falling in between rather than exceeding Audio. This suggests that adding text does not straightforwardly improve consensus once audio is already available for humans. Since different annotators are assigned to different modalities and the sample size is modest, the lower Audio+Text agreement relative to Audio should be interpreted cautiously; one possible explanation is that humans are more sensitive to conflicts between lexical/ASR-based cues in transcripts and acoustic impressions, and this heightened sensitivity increases uncertainty rather than simply adding information. Across classification granularity, agreement is consistently lower for multi-class than binary labels within every modality, reflecting the added difficulty of distinguishing fine-grained relationship subtypes versus the coarser familiar/stranger distinction. Taken together, these results suggest that this task is a difficult, high-disagreement task.

\subsection{Human Evaluation Results}
In the Text and Audio+Text modalities, the best methods outperform averaged individual human votes on all metrics, although not every model variant does so (see Table~\ref{tab:modality-results}), indicating that at the level of single annotator decisions, text-included relationship inference is highly uncertain. However, this should be read together with the low inter-annotator agreement in Table~\ref{tab:iaa}: when annotators disagree substantially, evaluating each individual vote against a single ground-truth label reflects not only annotator error but also the ambiguity of the task and the under-specification of relationship cues in transcripts. The Audio Modality shows a slightly different pattern: humans achieve higher binary macro-F1 (Table~\ref{tab:modality-results}), suggesting that human annotators are less biased toward the majority \textit{familiar} class and make more balanced binary decisions when acoustic cues are the only clues. The higher agreement in audio-based settings further suggests that prosodic and paralinguistic information—such as tone, laughter, hesitation, pacing, and turn-taking rhythm—provides shared perceptual evidence not fully captured by current model systems. Overall, the human evaluation could be read as a reference point rather than a strict upper bound. Models can outperform humans in some settings, while human annotators show stronger balanced binary performance in the audio modality. This suggests that current models exploit textual cues effectively, but may still be less capable than humans at leveraging acoustic evidence.

\section{Limitations}
This work has several limitations. First, experiments are conducted on a single benchmark, so additional datasets are needed to assess generalization. Second, the proposed multi-agent methods require multiple LLM calls and therefore increase inference-time cost; future work should study accuracy--cost trade-offs and more efficient agent interaction. Finally, several relationship categories are represented by relatively few test samples, so per-category performance should be interpreted with appropriate caution.

\section{Conclusions}

In this work, we studied speaker relationship inference as a 
social reasoning task and a step toward socially aware speech understanding. Unlike conventional dialogue tasks focused on intent, topic, or sentiment, relationship inference requires 
identifying latent interpersonal structure from open-ended interactions, where cues are subtle, distributed across turns, and expressed through both lexical and acoustic signals.
To address this challenge, we explored structured multi-agent reasoning as a training-free paradigm, proposing two inference-time frameworks: Multi-Role Multi-Agent Debate, which assigns agents complementary analytic roles, and Multi-Agent Compete, which competes through adversarial comparison. Across text, audio, and audio-text settings, these structured interactions improve, in most cases, over zero-shot prediction and prior debate-based baselines. This suggests that the benefit comes not merely from using multiple agents, but from designing interactions that better capture complementary and ambiguous relational cues, and more broadly, that
structured inference-time reasoning provides a practical alternative to task-specific training for socially grounded speech understanding. We hope this work motivates future research on applying structured LLM reasoning to other challenging conversational inference tasks involving latent social attributes, interaction dynamics, and multimodal conversational understanding.

\section{Acknowledgments}
A generative AI language model (ChatGPT) was used in the preparation of this manuscript to assist with grammar checking and language refinement. This assistance was limited to language-level editing and did not extend to the scientific content of the work. No ideas, hypotheses, experimental designs, implementations, code, datasets, analyses, or bibliographic references were generated by any AI system. All technical content and the final text were reviewed by the authors.

\bibliographystyle{plain}
\bibliography{references}

@article{agrawal2025seamless,
  title={Seamless interaction: Dyadic audiovisual motion modeling and large-scale dataset},
  author={Agrawal, Vasu and Akinyemi, Akinniyi and Alvero, Kathryn and Behrooz, Morteza and Buffalini, Julia and Carlucci, Fabio Maria and Chen, Joy and Chen, Junming and Chen, Zhang and Cheng, Shiyang and others},
  journal={arXiv preprint arXiv:2506.22554},
  year={2025}
}

@inproceedings{tigunova2021pride,
  title={PRIDE: Predicting relationships in conversations},
  author={Tigunova, Anna and Mirza, Paramita and Yates, Andrew and Weikum, Gerhard},
  booktitle={Proceedings of the 2021 conference on empirical methods in natural language processing},
  pages={4636--4650},
  year={2021}
}

@inproceedings{stark2012hello,
  title={Hello, who is calling?: can words reveal the social nature of conversations?},
  author={Stark, Anthony and Shafran, Izhak and Kaye, Jeffrey},
  booktitle={Proceedings of the 2012 conference of the North American chapter of the association for computational linguistics: human language technologies},
  pages={112--119},
  year={2012}
}

@inproceedings{yella2014inferring,
  title={Inferring social relationships in a phone call from a single party's speech},
  author={Yella, Sree Harsha and Anguera, Xavier and Luque, Jordi},
  booktitle={2014 IEEE International Conference on Acoustics, Speech and Signal Processing (ICASSP)},
  pages={4843--4847},
  year={2014},
  organization={IEEE}
}

@inproceedings{katerenchuk2014your,
  title={" was that your mother on the phone?": classifying interpersonal relationships between dialog participants with lexical and acoustic properties.},
  author={Katerenchuk, Denys and Brizan, David Guy and Rosenberg, Andrew},
  booktitle={INTERSPEECH},
  pages={1831--1835},
  year={2014}
}

@inproceedings{du2024improving,
  title={Improving factuality and reasoning in language models through multiagent debate},
  author={Du, Yilun and Li, Shuang and Torralba, Antonio and Tenenbaum, Joshua B and Mordatch, Igor},
  booktitle={Forty-first international conference on machine learning},
  year={2024}
}

@inproceedings{wu2025identifying,
  title={Identifying power relations in conversations using multi-agent social reasoning},
  author={Wu, Zhaoqing and Goldwasser, Dan and Pacheco, Maria Leonor and Morgenstern, Leora},
  booktitle={Proceedings of the 2025 Conference of the Nations of the Americas Chapter of the Association for Computational Linguistics: Human Language Technologies (Volume 2: Short Papers)},
  pages={855--865},
  year={2025}
}

@inproceedings{yang2026mad,
  title={MAD-Logic: Multi-Agent Debate Enhances Symbolic Translation and Reasoning},
  author={Yang, Haocheng and Cheng, Fengxiang and Yao, Tianjun and Yang, Mengyue and Chai, Jiajun and Wang, Xiaohan and Yin, Guojun and Lin, Wei and Kar, Soummya and Liu, Fenrong and others},
  booktitle={The Fourteenth International Conference on Learning Representations},
  year={2026}
}

@inproceedings{sun2025cortexdebate,
  title={CortexDebate: Debating Sparsely and Equally for Multi-Agent Debate},
  author={Sun, Yiliu and Zhao, Zicheng and Wan, Sheng and Gong, Chen},
  booktitle={Findings of the Association for Computational Linguistics: ACL 2025},
  pages={9503--9523},
  year={2025}
}

@article{yang2025socially,
  title={Socially aware language technologies: Perspectives and practices},
  author={Yang, Diyi and Hovy, Dirk and Jurgens, David and Plank, Barbara},
  journal={Computational Linguistics},
  volume={51},
  pages={689--703},
  year={2025}
}

@article{clark1979interpersonal,
  title={Interpersonal attraction in exchange and communal relationships.},
  author={Clark, Margaret S and Mills, Judson},
  journal={Journal of personality and social psychology},
  volume={37},
  number={1},
  pages={12},
  year={1979},
  publisher={American Psychological Association}
}

@article{pincus2003interpersonal,
  title={Interpersonal theory of personality},
  author={Pincus, Aaron L and Ansell, Emily B},
  journal={Handbook of psychology: Personality and social psychology},
  volume={5},
  pages={209--229},
  year={2003}
}

@article{wright2023contemporary,
  title={Contemporary integrative interpersonal theory: Integrating structure, dynamics, temporal scale, and levels of analysis.},
  author={Wright, Aidan GC and Pincus, Aaron L and Hopwood, Christopher J},
  journal={Journal of Psychopathology and Clinical Science},
  volume={132},
  number={3},
  pages={263},
  year={2023},
  publisher={American Psychological Association}
}

@article{fiske1990relativity,
  title={Relativity within Moose (" Mossi") culture: Four incommensurable models for social relationships},
  author={Fiske, Alan Page},
  journal={Ethos},
  volume={18},
  number={2},
  pages={180--204},
  year={1990},
  publisher={JSTOR}
}

@article{fiske1992four,
  title={The four elementary forms of sociality: framework for a unified theory of social relations.},
  author={Fiske, Alan P},
  journal={Psychological review},
  volume={99},
  number={4},
  pages={689},
  year={1992},
  publisher={American Psychological Association}
}

@book{krippendorff2018content,
  title={Content analysis: An introduction to its methodology},
  author={Krippendorff, Klaus},
  year={2018},
  publisher={Sage publications}
}

\end{document}